# Virtualizing System and Ordinary Services in Windows-based OS-Level Virtual Machines

Zhiyong Shan [§][*]
zshan@cs.sunysb.edu

Tzi-cker Chiueh[*]
chiueh@cs.sunysb.edu

Xin Wang[*]
xwang@ece.sunysb.edu

[§]Key Laboratory of Data Engineering and Knowledge Engineering, MOE, Renmin University of China

[*]Stony Brook University

**ABSTRACT**

OS-level virtualization incurs smaller start-up and run-time overhead than HAL-based virtualization and thus forms an important building block for developing fault-tolerant and intrusion-tolerant applications. A complete implementation of OS-level virtualization on the Windows platform requires virtualization of Windows services, such as system services like the Remote Procedure Call Server Service (RPCSS), because they are essentially extensions of the kernel. As Windows system services work very differently from their counterparts on UNIX-style OS, i.e., daemons, and many of their implementation details are proprietary, virtualizing Windows system services turned out to be the most challenging technical barrier for OS-level virtualization for the Windows platform. In this paper, we describe a general technique to virtualize Windows services, and demonstrate its effectiveness by applying it to successfully virtualize a set of important Windows system services and ordinary services on different versions of Windows OS, including RPCSS, DcomLaunch, IIS service group, Tlntsvr, MySQL, Apache2.2, CiSvc, ImapiService, etc.

**Categories and Subject Descriptors** D.4.5 [**Operating Systems**]:Reliability; D.4.6 [**Operating Systems**]: Security and Protection

**General Terms** Reliability, Security

## 1. INTRODUCTION

OS-level virtualization provides an excellent platform for the development of fault-tolerant and intrusion-tolerant applications, because, it incurs little or no overhead when creating, running, and shutting down a virtual machine (VM). OS-level VMs (i.e., containers) are designed to share as many resources as they can with other VMs or with the host environment. Moreover, programs in an OS-level VM run as normal applications that directly use the host operating system's system call interface and do not need to run on top of an intermediate hypervisor, which could be a specially crafted software (as in the case of VMware ESX [1] and Xen [3]) or a standard operating system (as in the case of Virtual PC [2] and UML [4]).

A standard implementation for OS-level virtualization is to intercept the system call interface, and rename the system resources being manipulated so that the system resources of each virtual machine reside in a separate name space. A well-known problem with OS-level virtualization is that all OS-level virtual machines running on top of a kernel share the kernel's state, because OS-level virtualization does not virtualize the kernel state. On the Windows platform, a set of user-level system services, which behave like daemons in a Unix-style OS, are used to augment the kernel and provide various critical functionalities to other services and applications. For example, Windows's inter-process communication mechanisms such as COM, DCOM and RPC, are supported by the RPCSS service. How to virtualize these system services so that the large install base of Windows applications can run correctly in different VMs without interfering one another turns out to be a major implementation challenge.

The goal of virtualizing Windows services is to duplicate a separate instance of each Windows service in each VM. However, it is non-trivial to completely achieve this goal for the following reasons. First, because some Windows system services, e.g., RPCSS, are really parts of the Windows OS, the kernel does not allow them to be duplicated. That is, Windows is designed to forbid users or applications to create, register and run multiple instances of the same system service simultaneously on a single OS. Second, to duplicate Windows services, one needs to intercept and change inter-process communications, inter-service co-operations, and registry manipulations, etc. However, because the details of these behaviors are largely proprietary, it is difficult to correctly reverse-engineer all of them and make necessary modifications to them. Third, in some cases, even correctly handling all relevant behaviors is not sufficient because some Windows services hard-code certain resource names in their binaries and therefore may break when they run on top of an OS-level virtualization layer that renames system resources.

In this paper, we propose a novel and general scheme to virtualize Windows services that consists of four steps: logically duplicating a Windows service in Service Control Manager (SCM)'s internal database, physically duplicating a Windows service by starting a new process and putting it into its associated VM, picking out and changing inter-service co-operations, identifying and modifying system resource names that are hard-coded into service binaries. To test the effectiveness of the proposed scheme, we applied it to successfully virtualize several important system services on different versions of Windows OS, including the RPCSS service, which plays a critical role in RPC/COM/DCOM functionalities. We believe our technique is the first that successfully demonstrates that multiple instances of the RPCSS service can run concurrently on top of the Windows OS. In addition, our technique allows multiple instances of the IIS or Apache web server, or Mysql database server to run simultaneously on a Windows machine.

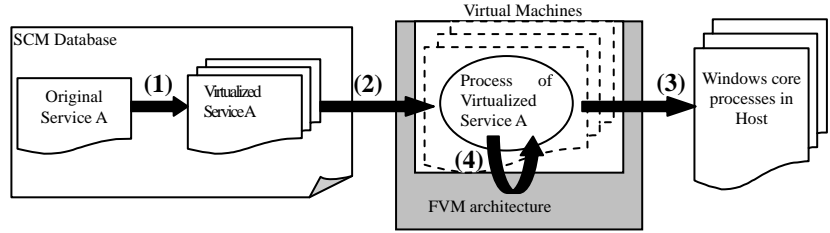

**Figure 1.** Windows service virtualization scheme consisting of four-step

The rest of this paper is organized as follows. We first present the proposed technique for Windows service virtualization in Section 2, and evaluate the functionality and performance of the prototype in Section 3. In Section 4, we compare this research with other similar efforts in the literature. In Section 5, we summarize the main contributions of this work and outline future research directions.

## 2. SOLUTION

We propose a generic scheme to virtualize Windows services. As depicted in Figure 1, it consists of the following four steps:

(1) Logically duplicating a Windows service by creating a separate entry in the SCM database with a new service name.

(2) Physically duplicating a new instance of a service by starting a service process and putting it in its corresponding VM.

(3) Managing the inter-service interactions between virtualized services and Windows core processes.

(4) Renaming hard-coded service names embedded in service binaries.

In the rest of this section, we describe these four steps in detail.

### 2.1 Creating a Virtualized Service

To create a new Windows service, the API function `CreateService()` is a natural choice. This API function, however, could not create a new instance of a DLL-based service just like the original instance, because `CreateService()` could neither set up a DLL-based service name in the SCM database nor add a new service group into a `svchost.exe` process. Another method is to modify the SCM database directly in order to add a new DLL-based service. However, this method could not start the new service immediately after the SCM database modification, because SCM cannot incorporate the SCM database modification until the whole system is rebooted.

In the end, we choose to combine these two methods to create a new Windows service as follows:

(1) Calling `CreateService()` with a virtualized service name, like "ServiceName-vmX", where ServiceName is the original name of the service to be virtualized and X is the VM ID of the VM initiating the service virtualization request, to inform SCM the creation of a new service. This way we can launch the newly created service without rebooting the entire system.

(2) Making a copy of the SCM database entries associated with the virtualized service, and modifying three places in the copied SCM database entries. First, the original service name should be changed to a virtualized service name, e.g., from "ServiceName" to "ServiceName-vmX". Second, the names of the services on which the virtualized service depends should also be changed from their original service names to virtualized service names. For example, because the IISADMIN service depends on RPCSS, when virtualizing the IISADMIN service, the service name of RPCSS should be changed to RPCSS-vmX. Last, the start type of the new virtualized service should be set to `manual start` rather than `automatic start`, because a virtualized service should be started after the boot-up of a VM. All other SCM database entries should be left untouched. With these SCM database entries, a virtualized service will behave exactly the same as the original service from which it is cloned.

### 2.2 Starting a Virtualized Service Process

Calling the API function `StartService()` causes SCM to launch a new service process. To determine which VM the newly launched service process should be placed in, we need a mechanism to determine to which VM the started service belongs.

At the time when a VM creates a service, we record a FVM flag and the VM's ID in the SCM database entry that contains the service's binary path and start-up parameter. When the SCM launches a service process, FVM's in-kernel monitor moves the service process into its corresponding VM according to the FVM flag and VM ID in the process' pathname and parameters.

**Table 1.** IPC objects facilitating interactions between virtualized services and Windows core processes

| IPC Type | IPC Objects |
|---|---|
| Port | \RPC Control\DNSResolver<br>\RPC Control\ntsvcs |
| Named Pipe | \Device\NamedPipe\net\NtControlPipe* (* represents an arbitrary number)<br>\Device\NamedPipe\svcctl (only on Windows 2k)<br>\Device\NamedPipe\ntsvcs<br>\Device\NamedPipe\EVENTLOG<br>\Device\NamedPipe\samr |
| Mutex | \BaseNamedObjects\DBWinMutex<br>\BaseNamedObjects\RasPbFile<br>\BaseNamedObjects\SHIMLIB_LOG_MUTEX<br>\BaseNamedObjects\ShimCacheMutex |
| Section | \BaseNamedObjects\__R_ 0000000000da_SMem__<br>\BaseNamedObjects\DBWIN_BUFFER<br>\BaseNamedObjects\ShimSharedMemory |
| Event | \BaseNamedObjects\ScmCreatedEvent<br>\BaseNamedObjects\SvcctrlStartEvent_A3752DX<br>\BaseNamedObjects\crypt32LogoffEvent<br>\BaseNamedObjects\userenv: User Profile setup event<br>\BaseNamedObjects\DINPUTWINMM<br>\SECURITY\LSA_AUTHENTICATION_INITIALIZED |

For a DLL-based service, SCM starts it as a thread inside a `svchost.exe` process. In this case we record a FVM flag and a VM ID at the end of the service process' parameter list in the SCM database. For example, the original RPCSS service has a parameter like "-k rpcss", the virtualized RPCSS service will have a parameter like "-k rpcss-vmX", "X" again represents a VM ID. When the new RPCSS service process is launched with parameter "-k rpcss-vmX" and FVM's in-kernel monitor sees a process with a start-up parameter containing "-vmX", it adds the process into the VM whose VM ID is X.

For an EXE-based service, SCM starts it as an independent process. We first copy the EXE file into the VM's workspace and then record its pathname in the SCM database. Because the pathname is in a VM, it will naturally contain a FVM flag and a VM ID. For example, `W3SVC` is a Windows service for web service, the pathname of the original W3SVC service's image is "`c:\WINNT\system32\inetsrv\inetinfo.exe`", and the pathname of its virtualized version becomes "`c:\fvms\VM-X\C\WINNT\system32\inetsrv\inetinfo.exe`", where "X" represents the ID of the VM in which the virtualized service is to be placed. When FVM's in-kernel monitor sees a process with such an image pathname, it will put the process in the corresponding VM.

## 2.3 Maintaining Existing Inter-Service Interactions

Most inter-service dependencies can be derived from SCM database entries, i.e. the `DependOnService` and `DependOnGroup` entries under a service's registry key. However, some inter-service dependencies are not explicitly recorded in the SCM database, and can only be identified through reverse engineering. For example, on the Windows XP platform the RPCSS service depends on the DcomLaunch service, because starting RPCSS requires a shared global data structure `RotHintTable` which is created by DcomLaunch. Once these inter-service dependencies are identified, we follow a simple principle to ensure these dependencies are observed in the service virtualization process:

*The starting order of the virtualized services in a VM should be the same as that of their original services in the host.*

The Windows core processes, e.g. SCM, Lsass and WinLogon, are closely related to the Windows kernel itself and therefore cannot be duplicated or virtualized, i.e., running a new instance in each VM. These core processes therefore run in the host and virtualized services interact with them through IPC objects. Under FVM, when a process in a VM interacts with other processes in the same VM using IPC, FVM's in-kernel monitor intercepts the IPC requests and renames the resources referred to in the parameters of the requests. However, for the IPC requests used by a virtualized service running in a VM to interact with the core processes which run in the host, the arguments used in the IPC requests should *not* be transformed. For instance, Windows services often connect to the SCM process through a named pipe called `NtControlPipe` during start-up by calling the API function `CreateFile()` with the file name `NtControlPipe`. When a virtualized service running in a VM tries to open the named pipe by calling the API function `CreateFile()`, if the file name argument were changed to `NtControlPipe-vmX` by FVM, the virtualized service would not be able to interact with SCM because SCM runs in the host and still uses the original pipe name `NtControlPipe`.

One way to fix this problem is to identify all IPCs used by virtualized services to interact with the core Windows processes and avoid renaming the resource arguments used in these IPC calls. However, it is very challenging to identify all such IPC calls, because the implementation details of Windows core processes are mostly undocumented. To identify long-term IPC objects and IPC objects used after process start-up, we statically analyze the shared IPC objects between Windows core processes and virtualized services using the tool `ProcessExplorer`. To pinpoint short-term IPC objects and IPC objects used during process start-up, we logged IPC-related system calls and certain Win32 API function calls during a service's execution in a VM and compared the resulting log with that associated with the same service's execution in the host. Eventually, we were able to figure out all the IPC objects used by Windows core processes on the Windows 2K and XP platforms, and list them in Table 1. With this IPC object list, when a Windows service running in a VM utilizes an IPC object, FVM will not rename the IPC objects involved if it is in the list, thus enabling the service to seamlessly interact with the core processes running in the host.

## 2.4 Renaming Hard-Coded Service Names

After the above three steps, FVM can correctly virtualize many Windows services. However, some services, after being virtualized, could not complete its start-up procedure, for instance, the RPCSS service. By disassembling these services' binary files, we found the root cause is the "hard-coded service name" problem. Figure 2 shows the disassembly result of several code fragments of the Windows XP version of `rpcss.dll`. In these code fragments two hard-coded RPCSS service name strings are used as input arguments to the string manipulation function `RtlInitUnicodeString()` and to the service management function `OpenServiceW()`. When the rpcss.dll is invoked in a VM with service name RPCSS–vmX, the hard-coded name will be sent to SCM by the function `OpenServiceW()`. If at the time another RPCSS service is running in a different VM, SCM will refuse the `OpenServiceW()` function since both services use the hard-coded name RPCSS. Eventually, the service RPCSS-vmX fails to complete its start-up procedure.

After analyzing a bunch of Windows service binaries, we found there are two types of API functions that use hard-coded service name as input arguments. The first type is service-related API functions and the other type is string manipulation-related API functions. Accordingly, we solve the "hard-coded service name" problem by intercepting these two types of API functions issued by a virtualized service, and checking if the associated argument is a virtualized service name or an unmodified service name. If it is an original service name, we change it to the corresponding virtualized service name, because functions in virtualized service processes should only use virtualized service names.

## 3. PROTOTYPE

To demonstrate the effectiveness of the proposed service virtualization scheme, we developed a prototype based on FVM on Windows 2K and XP. Although XP and 2K are not new, they are enough for verifying the approach of service virtualization since both versions of Windows OS have very similar system calls, Win32 API functions and service mechanisms, based on which the virtualization layer works.

To verify the functionality of our prototype, we successfully virtualized several important Windows services using the prototype. On Windows 2k, we virtualized RPCSS and IIS service groups. On Windows XP, we virtualized RPCSS service group, MySQL service for Mysql database, Apache2.2 service for Apache web server, Tlntsvr service for telnet server, CiSvc service for indexing files, ImapiService server for managing CD recording, etc. The virtualized services work smoothly inside VMs.

For example, the virtualization steps finished by the prototype for the RPCSS service are as follows (suppose the VM ID of the VM that requests a new instance of RPCSS is Z):

(1) Create the registry key `HKLM\SYSTEM\CurrentControlSet\Services\RpcSs-vmZ` and the registry value `RpcSs-vmZ` under `HKLM\SOFTWARE\Microsoft\WindowsNT\CurrentVersion\SvcHost`.

(2) Start RPCSS service and recognize its process by the parameter "`-k rpcss-vmZ`", then move the process into VM Z.

(3) Enable the interaction between RPCSS and the core processes running in the host by disabling resource renaming for accessing IPC objects listed in Table 1.

(4) The hard-coded service name `rpcss` in the service binary `rpcss.dll` is used by a service-related API function `RegisterServiceCtrlHandlerEx()` as an argument. We rename it to `rpcss-vmZ`.

Because the success of RPCSS virtualization, we now can run several applications that need the services provided by

```
76A9752F  push        esi
76A97530  push        00000004h
76A97532  push        SWC76A975C4_RpcSS
76A97537  push        [L76ABE8B4]
76A9753D  call        [ADVAPI32.dll!OpenServiceW]
……
76A975C4  SWC76A975C4_RpcSS:
76A975C4  unicode 'RpcSS',0000h
……
76A97C0C  SWC76A97C0C_RPCSS:
76A97C0C  Unicode 'RPCSS',0000h
……
76A989FF  mov         edi,[ebp+0Ch]
76A98A02  mov         [L76ABE390],esi
76A98A08  push        [edi]
76A98A0A  mov         esi,[ntdll.dll!RtlInitUnicodeString]
76A98A10  lea         eax,[ebp-18h]
76A98A13  push        eax
76A98A14  mov         [ebp-08h],ebx
76A98A17  call        esi
76A98A19  push        SWC76A97C0C_RPCSS
76A98A1E  lea         eax,[ebp-10h]
76A98A21  push        eax
76A98A22  call        esi
76A98A24  push        00000001h
76A98A26  lea         eax,[ebp-10h]
76A98A29  push        eax
76A98A2A  lea         eax,[ebp-18h]
76A98A2D  push        eax
76A98A2E  call        [ntdll.dll!RtlEqualUnicodeString]
……
76AA8EC4  call        [ADVAPI32.dll!OpenSCManagerW]
76AA8ECA  cmp         eax,edi
76AA8ECC  mov         [ebp-04h],eax
76AA8ECF  jz          L76AA8F16
76AA8ED1  push        ebx
76AA8ED2  push        esi
76AA8ED3  push        000F01FFh
76AA8ED8  push        SWC76A97C0C_RPCSS
76AA8EDD  push        eax
76AA8EDE  call        [ADVAPI32.dll!OpenServiceW]
```

**Figure 2.** Code fragments of anti-compiled rpcss.dll. Two hard-coded RPCSS service name strings are used as input arguments to functions RtlInitUnicodeString() and OpenServiceW().

RPCSS, including Microsoft Office Assistant, Excel COM server, which is launched when users try to edit an Excel object inside a Word document, Adobe installation program, etc.

.As the performance overhead of service virtualization is mainly resulted from executing additional instructions when intercepting system calls and API functions, we measure specifically the interception overhead of the corresponding system calls and API functions. The per-system-call or per-API-function-call overhead of the proposed service virtualization scheme is small to negligible when compared with the baseline Windows OS, and its penalty on the startup time and the run-time performance of virtualized services is much smaller than those when the same service is virtualized on top of VMware Workstation. This performance difference mainly comes from the fact that FVM is designed to enforce inter-VM isolation through logical name renaming, whereas VMware Workstation is designed to enforce inter-VM isolation through physical resource separation.

## 4. RELATED WORK

As far as we know, there is no such a project successfully virtualized both ordinary service and system service on Windows OS in literature. There are two projects more closely related to our work. One is Feather-weight Virtual Machine (FVM) [5] that enables multiple isolated execution environments to run on a single Windows kernel. It is able to virtualize a limited number of ordinary services, whereas, it can not virtualize system services such as RPCSS and ordinary services that require complex inter-service interactions, e.g. IIS service group. The other project is Virtuozzo [8] that provides isolated environments called Virtual Dedicated Server or Virtual Private Server (VPS) on Windows platform, but we could not find any description about Windows service virtualization, especially system service virtualization from its public document.

The original version of FVM can only virtualize a limited number of Windows services by intercepting service-related API functions and attaching FVM flags and VM identification numbers to service names. However, as explained in Section 3.1, this method is not only limited but also incorrect in some cases because not all Windows services can be duplicated. For instance, the RPCSS service runs as a thread inside a svchost.exe process and therefore cannot be duplicated by just calling the standard service API function. Furthermore, because the original version of FVM did not recognize let alone address the "hard-coded service name" problem, it can not handle services whose binary contains hard-coded service names.

Other commercial products on Windows also include similar OS-level virtualization techniques, including Softricity Desktop [9], AppStream [10] and Thinstall [11]. In particular, Softricity Desktop [9] implements comprehensive virtualization to execute virtualized applications without requiring any pre-installation. Specifically it enables each virtualized application to execute against its own set of registries and configuration files within a virtual machine on any machine to which the virtualized application is deployed. However, Softricity Desktop can only provide isolated runtime environments for applications but not for Windows services.

## 5. CONCLUSION

OS-level virtualization virtualizes the system resource at the system call interface, and relies on a single kernel to provide system resources to multiple virtual machines running on an OS-level virtualization layer. On the Windows platform, parts of the kernel's functionalities are actually embodied in some user-level services. As a result, these services are tightly integrated with the kernel and therefore cannot be duplicated or virtualized in each VM. However, it is still essential to virtualize as many Windows services as possible so that OS-level virtualization can become a viable virtualization technology on Windows. Unfortunately, none of the publicly available documents on OS-level virtualization technologies ever mention the service virtualization problems, not to mention addressing them. We thus believe this paper is the first to describe a generic Windows service virtualization scheme that can virtualize not only ordinary services, but also important system services such as RPCSS. Applying this scheme, we successfully virtualized RPCSS, DcomLaunch and IIS service group including IISADMIN, W3SVC, MSFTPSVC, SMTPSVC and NNTPSVC on both Windows 2K and XP, as well as MySQL, Apache2.2, CiSvc, ImapiService and Tlntsvr on Windows XP. As a result, we can successfully run multiple instances of the IIS web server, Apache web server or MySQL database server simultaneously on a single Windows machine. Empirical performance measurements on the prototype implementation of the proposed service virtualization scheme show that the additional performance overhead introduced by service virtualization is rather minor when compared with the overhead introduced by the original version of FVM. Moreover, the startup time and run-time performance of virtualized services of FVM with service virtualization are about two to three times better than the same set-ups running on VMware Workstation. These results demonstrate the potential performance advantage and thus importance of the proposed service virtualization technique.

## 6. ACKNOWLEDGMENTS

This work has been supported by the National Science Foundation of China (NSFC) under grants 60703103 and 60833005.